\documentclass[12pt]{article}

\usepackage{graphicx}
\usepackage{amsfonts}
\def\+{{+\!\!\!+}}

\def\d{\partial}

\def\N{\nabla}

\def\l{\lambda} 
 
\def\L{\Lambda}

\def\F{\Psi} 
\def\e{\varepsilon}

\def\ka{K\"ahler\ }

\def\pmb#1{\setbox0=\hbox{#1}%
\kern.0em\copy0\kern-\wd0 
\kern-.04em\copy0\kern-\wd0 
\kern.08em\copy0\kern-\wd0 
\kern-.04em\raise.0433em\box0 }         
 
\def\half{\textstyle{1\over 2}}

\newcommand{\nc}{\newcommand} 
\nc{\beq}{\begin{equation}} 
\nc{\eeq}[1]{\label{#1}\end{equation}} 
\nc{\ber}{\begin{eqnarray}} 
\nc{\eer}[1]{\label{#1}\end{eqnarray}} 
\nc{\pek}[1]{\cite{#1}} 
\nc{\enr}[1]{(\ref{#1})} 
\nc{\kal}[1]{{\cal{#1}}} 
\nc{\dott}{\;\cdot\;} 

\def\0 {\nonumber}

\begin{document} 
\setcounter{page}{0}
\newcommand{\inv}[1]{{#1}^{-1}} 
\renewcommand{\theequation}{\thesection.\arabic{equation}} 
\newcommand{\be}{\begin{equation}} 
\newcommand{\ee}{\end{equation}} 
\newcommand{\bea}{\begin{eqnarray}} 
\newcommand{\eea}{\end{eqnarray}} 
\newcommand{\re}[1]{(\ref{#1})} 
\newcommand{\qv}{\quad ,} 
\newcommand{\qp}{\quad .} 
\begin{titlepage} 
\begin{center} 
                          
\hfill   hep-th/0411186\\ 
                          \hfill   UUITP-25/04\\
                          \hfill   HIP-2004-64/TH\\
                          \hfill   YITP-SB-04-63\\

\vskip .3in \noindent 


{\large \bf{Generalized \ka geometry and manifest $\kal{N}=(2,2)$ supersymmetric nonlinear sigma-models}} \\

\vskip .2in 

{\bf Ulf~Lindstr\"om$^{ab}$,} {\bf Martin Ro\v cek$^{c}$,}
{\bf Rikard von Unge$^{d}$,}
and {\bf Maxim Zabzine$^{e}$,}

\vskip .05in 
$^a${\em\small Department of Theoretical Physics\\ 
Uppsala University, Box 803, SE-751 08 Uppsala, Sweden} \\
\vskip .02in
$^b${\em\small HIP-Helsinki Institute of Physics\\
P.O. Box 64 FIN-00014 University of Helsinki, Suomi-Finland}\\
\vskip .05in 
$^c${\em C.N.Yang Institute for Theoretical Physics, SUNY, \\
Stony Brook, NY 11794-3840,USA}
\vskip .05in 
$^{d}${\em Institute for Theoretical Physics, Masaryk University, 61137 Brno, Czech Republic }
\vskip .05in
$^e${\em School of Mathematical Sciences, Queen Mary, University of London, \\
Mile End Road, London, E1 4NS, UK}
\vskip .5in 
\end{center} 
\begin{center} {\bf ABSTRACT }  
\end{center} 
\begin{quotation}\noindent  
Generalized complex geometry is a new mathematical framework that is useful for 
describing the target space of $\kal{N}=(2,2)$ nonlinear sigma-models.
The most direct relation is obtained at the  $\kal{N}=(1,1)$ 
level when the sigma model is formulated with an additional 
auxiliary spinorial field. We revive a formulation in terms of  
$\kal{N}=(2,2)$ semi-(anti)chiral multiplets where such auxiliary fields are naturally present. 
The underlying generalized complex structures are shown to commute 
(unlike the corresponding ordinary complex structures) and describe a 
Generalized \ka geometry. The metric, $B$-field and generalized complex 
structures are all determined in terms of a potential $K$.
\end{quotation} 
\vfill 
\eject 


\end{titlepage}

\section{Introduction}
In this paper, we show how a sigma-model based on 
semi-chiral and semi-antichiral superfields realizes a new mathematical 
concept called a generalized complex structure (GCS).

The geometry of the target space of supersymmetric nonlinear sigma-models is restricted by the
number of supersymmetries in the base as well as the type of background fields (metric $g$ and
$NS-NS$ two-form $B$) in the target space \cite{Zumino:1979et}, \cite{Alvarez-Gaume:1980vs}, 
\cite{Gates:1983py}, \cite{Gates:nk}.  The situation in two dimensions is summarized in Table 1:
\vskip .1in
\begin{table}[htbp]
\centering
\begin{tabular}{|l|c|c|c|c|c|}
\hline
Supersymmetry & (0,0) or (1,1) & (2,2) & (2,2) & (4,4) & (4,4) \\ 
\hline
Background & $g,B$ &  $g$ & $g,B$ & $g$ & $g,B$ \\
\hline
Geometry   &  Riemannian & K\"ahler & bihermitian & hyperk\"ahler & bihypercomplex \\
\hline
\end{tabular}
\caption{The geometries of sigma-models with different supersymmetries.}
\end{table}
\vskip -.1in
\noindent Some intermediate $\kal{N} =(p,q)$ geometries are also partly classified  
\cite{AbouZeid:1996za},  \cite{Abou-Zeid:1999em}.

We focus on  $\kal{N} =(2,2)$ models in a metric and a $B$-field  background, and find a 
reinterpretation of the bihermitean geometry. This is  motivated  
by some recent advances in mathematics, initiated by Hitchin to describe generalized Calabi-Yau 
manifolds, {\it e.g.,}  including an antisymmetric $B$-field \cite{Hitchin}. This Generalized Complex Geometry (GCG) \cite{Gualtieri} includes bihermitean geometry, \ka geometry, and symplectic geometry as  special cases. Its fundamental object is a Generalized 
Complex Structure (GCS), which is a map of the sum of the tangent and cotangent spaces  $T\oplus T^*$ to itself; the GCS squares to minus one and obeys an integrability
condition stated in terms of the Courant bracket, a generalization of the Lie bracket to $T\oplus T^*$. The GCG is very well suited to a description of  $\kal{N} =(2,2)$ in a nontrivial general background for many reasons, one of which is that the automorphism group of the Courant bracket includes
the $b$-transform, which is precisely the gauge transformation of the $B$-field. 

The realization of GCG in supersymmetric nonlinear sigma-models has recently 
been investigated in \cite{Lindstrom:2004eh}, (a preliminary study of 
the  $\kal{N} =(2,2)$ model) and in \cite{Lindstrom:2004iw}, where the 
relation to GCS was established for  the $\kal{N} =(2,0)$ model. The full
$\kal{N} =(2,2)$ model in terms of its most general $\kal{N} =(1,1)$ 
formulation is not yet fully understood. Here we investigate how the 
GCS arises in models based on semi-(anti)chiral multiplets \cite{Buscher:1987uw}. 
These models have an underlying bihermitean geometry with noncommuting 
complex structures and a closed off-shell supersymmetry algebra. 
This sets them apart from the other known models with bihermitean geometry
and off-shell closure, {\it i.e.}, those formulated in terms of chiral and twisted chiral multiplets, which
have complex structures that necessarily commute. 

In \cite{Gualtieri} the concept of  generalized \ka geometry is introduced and shown to encode the bihermitean geometry of \cite{Gates:nk} into two {\em commuting}
GCS's subject to certain conditions.  Since the sigma model we discuss 
reduces to that of \cite{Gates:nk} when auxiliary fields are integrated out, we expect to find
left and right GCS's that together describe a generalized \ka geometry. There are many possible sets of auxiliary fields, however, and it is not a priori clear that the
ones we choose are the correct ``coordinates'' for such a description.

For a ``toy model'' based on semi-chiral multiplets only, we discover that the underlying geometry is indeed a generalized \ka geometry. 
This model is not a sigma model proper; nevertheless
the second supersymmetry transformations are determined in terms of the GCS's
corresponding to the complex structure and to the \ka form.

For the full model based on semi-chiral and semi-antichiral multiplets we uncover two GCS's corresponding to a second  left and a second  right supersymmetry.
Although the corresponding complex structures do not commute, the GCS's do. Our model thus satisfies part of the requirement for a generalized \ka geometry,
and we subsequently verify that the metric on $T\oplus T^*$ formed as the product of the two GCS's, is idempotent. The underlying GCG is thus the expected generalized \ka  geometry.  Further,  the metric, the $B$-field, and the GCS's are all determined by one and the same potential $K$.

The literature on GCG in physics is not yet very extensive. The geometry has been discussed in connection to generalized Calabi-Yau geometries in \cite{Grana:2004bg}, a short review of the 
sigma-model application appeared in \cite{Lindstrom:2004cd}, applications to  topological sigma-models was published in \cite{Kapustin:2004gv}, \cite{Chiantese:2004pe}
and other sigma model applications were discussed in \cite{Zucchini:2004ta}, \cite{Bergamin:2004sk}.

The organization of our presentation is as follows: In section two we introduce supersymmetric nonlinear sigma-models and recapitulate the features of bihermitean geometry. Section three contains a brief introduction to generalized complex geometry and section four contains an even briefer introduction to sigma model realizations of GCS. In section five we introduce the basics of  semi-(anti)chiral multiplets and section six contains various actions for them. Section seven shows how the 
GCS's emerge for the models discussed in section six. Section eight, finally, contains our conclusions and an outlook.

\section{Sigma models}
A nonlinear sigma model is a theory of maps
\ber
&&X^\mu(\xi):\kal{M}\rightarrow \kal{T}~,
\eer{map}
where $\xi^{i}$ are coordinates on $\kal{M}$ and $X^\mu$ coordinates on the target space 
$\kal{T}$. Classical solutions are found by extremizing the action.
\ber
&&S=\int d\xi ~\partial_{i}X^\mu \, g_{\mu\nu}(X)\,\partial^{i}X^\nu~,
\eer{act}
where the symmetric tensor $g_{\mu\nu}$ is identified with a metric on $\kal{T}$. 
The corresponding geometry is Riemannian for the bosonic model, but becomes complex when we impose enough supersymmetry. 

Supersymmetry is introduced by replacing the $X^\mu$'s by superfields:
\ber
&&X^\mu(\xi)\to\phi^\mu(\xi,\theta)~.
\eer{suf}
We shall be interested in a two dimensional $\kal{M}$, the dimension relevant for string theory.
There are (at least) two special features in $D=2$. First, there can be different amounts of supersymmetry in the left and right moving sectors denoted $\kal{N}=(p,q)$ supersymmetry. Second, if parity breaking terms are allowed,
the background may contain an antisymmetric $B_{\mu\nu}$-field.
For $\kal{N}=(2,2)$, the supersymmetric action written in terms of real $\kal{N}=(1,1)$ superfields  reads (with spinorial indices $\alpha = \{+,-\}$)
\ber
S=\int d^2\xi \, d^2\theta\, D_{+}\phi^\mu E_{\mu\nu}(\phi)D_{-}\phi^\nu~,
\eer{sac2}
where $E_{\mu\nu}(\phi)\equiv g_{\mu\nu}(\phi)+B_{\mu\nu}(\phi)$. This action has manifest 
$\kal{N}=(1,1)$ supersymmetry without any additional restrictions on the target space geometry. 
In \cite{Gates:nk}, Gates, Hull and Ro\v cek showed that it has additional nonmanifest supersymmetries,
\ber
&&\delta\phi^\mu= \e^+D_{+}\phi^\nu \mathbb{J}_{\mu}^{(+)\nu}+\e^-D_{-}\phi^\nu \mathbb{J}_{\mu}^{(-)\nu}~,
\eer{susy2}
(where $\{(+),(-)\}$ are labels and {\em not} spinor indices)
provided that the following conditions are fulfilled
\bigskip

\noindent
$\bullet $ Both the $J$'s are {\em almost complex structures}, {\it i.e.}, $\mathbb{J}^{(\pm)2}=-1$.\\
$\bullet $ They are {\em integrable}, {\it i.e.}, their Nijenhuis tensors vanish
\beq
\kal{N}_{\mu\nu}^{(\pm)\rho}\equiv  \mathbb{J}_{\mu}^{(\pm)\l}\d_{[\l} 
\mathbb{J}_{\nu]}^{(\pm)\rho}-(\mu\leftrightarrow\nu) =0
\eeq{nij}
$\bullet $  The metric is {\em hermitean} with respect to both complex structures, {\it i.e.}, it is preserved
by both structures:
$\qquad\qquad~ \mathbb{J}^{(\pm)t}g\mathbb{J}^{(\pm)}=g$\\
$\bullet $ The $\mathbb{J}$'s are {\em covariantly constant} with respect to a torsionful connection:
$$\N^{(\pm)}\mathbb{J}^{(\pm)}=0~,$$
where $\N^{(\pm)}\equiv \N ^0\pm g^{-1}H$ is the Levi-Civita connection plus(minus) the completely 
antisymmetric torsion given by the field-strength $H=dB$. 
(There are alternative, equivalent descriptions; see, {\it e.g.,} \cite{Lyakhovich:2002kc}.)

The above conditions represent a bihermitean target space geometry with a $B$-field, and result from 
requiring invariance of the action (\ref{sac2}) under the transformations (\ref{susy2}) as well as closure of 
the algebra of these transformations. Closure is only achieved on-shell, however. Only under the special 
condition that the two complex structures commute does the algebra close off-shell. In that case there is a manifestly $\kal{N}=(2,2)$ action for the model, given in terms of chiral and twisted chiral $\kal{N}=(2,2)$ superfields \cite{Gates:nk}.

If one is willing to introduce additional (auxiliary) spinorial $\kal{N}=(1,1)$ superfields, 
it is known how to accomodate noncommutativity of the complex structures for a special 
case which also has a manifest $\kal{N}=(2,2)$ formulation. 
This case corresponds to the semi-(anti)chiral 
superfields \cite{Buscher:1987uw} that we discuss below.

An interesting question is thus: What is the most general $\kal{N}=(2,2)$ sigma model with off-shell 
closure of the algebra, and what is the corresponding geometry? In asking this we  have in mind an 
extension of the model similar to the semi-chiral models, {\it i.e.},  to include additional fields to allow 
off-shell closure in the usual ``auxiliary field'' pattern and a geometry that includes these fields.

The GCG does contain the bihermitean geometry as a special case and thus seems a promising 
candidate. We therefore turn to a brief description of the GCG.

\section{Generalized Complex Geometry}

To  understand the generalization, let us first briefly look at some aspects of the definition of the 
ordinary complex structure. The features we need are that an almost complex structure 
$J$ on a $d$-dimensional manifold $\kal{T}$ is a map from the tangent bundle $J: T\to T$ 
that squares to minus the identity $J^2=-1$. With these properties 
$\pi_{\pm}\equiv \frac 1 2 (1\pm iJ)$ are projection operators, and we 
may ask when they define integrable distributions. The condition for this is that
\beq
\pi_{\mp}[\pi_{\pm}X,\pi_{\pm}Y]=0
\eeq{int}
for $X,Y\in T$ and $[,]$ the usual Lie-bracket on $T$. This relation is equivalent to the vanishing of 
the Nijenhuis tensor $\kal{N}(J)$, as defined in (\ref{nij}).

To define GCG, we turn our attention from the tangent bundle $T(\kal{T})$ to the 
sum of the tangent bundle and the co-tangent 
bundle $T\oplus T^*$. (Note that the structure group of this bundle is $SO(d,d)$, the string theory
T-duality group\footnote{This connection between  $T\oplus T^*$ was early on made in \cite{Duff:1990hn}}).
We write an element of $T\oplus T^*$ as $X+\xi$ with the vector $X\in T$ and the one-form
$\xi\in T^*$. The natural pairing $(X+\xi ,X+\xi)=\imath _{X}\xi$
gives a metric $\kal{I}$ on 
$T\oplus T^*$ as $X+\xi$, which in a coordinate basis $(\d_{\mu} ,dx^\nu)$ reads
\beq
\kal{I}=\left(\begin{array}{cc}
0&1_{d}\cr
1_{d}&0\end{array}\right)~.
\eeq{Imet}
In the definition of a complex structure above we made use of the Lie-bracket on $T$. To define 
a generalised complex structure we will need a bracket on $T\oplus T^*$. The relevant bracket is the 
the skew-symmetric {\em Courant bracket} \cite{courant} defined by\footnote{It 
does not in general satisfy the Jacobi 
identity; had it satisfied the Jacobi 
identity $(T\oplus T^*, [,]_{c})$ would have formed a Lie algebroid. It {\em does} satisfy
the Jacobi identity on subbundles  $L\subset T\oplus T^* $ that are Courant involutive and
isotropic with respect to $\kal{I}$ that is, subbundles that close under the bracket and whose sections 
are null with respect the metric $\kal{I}$ (for details see chapter 3 in \cite{Gualtieri}), 
but fails to do so in general. It fails in an interesting way which leads to the definition of a Courant algebroid \cite{Gualtieri}.

Another physical context where the Courant bracket 
naturally arises is that of anomaly-freedom of generalized currents recently discussed in \cite{Alekseev:2004np}.}  
\beq
[X+\xi , Y+\eta]_{c}\equiv [X,Y]+\pounds_{X}\eta-\pounds_{Y}\xi
-\half d(\imath _{X}\eta -\imath _{Y}\xi)~.
\eeq{cour}
This bracket equals the Lie-bracket on $T$ and vanishes on $T^*$. 
The most important property for us in the context of 
sigma-models is that its group of automorphisms is not only $Diff(\kal{T})$ but also
{\em $b$-transforms} defined by closed two-forms $b$,
\beq
e^b(X+\xi)\equiv X+\xi+\imath _{X}b~,
\eeq{btf}
namely,
\beq
[e^b(X+\xi), e^b(Y+\eta)]_{c}= e^b[X+\xi ,Y+\eta]_{c}~.
\eeq{btf2}

A {\em generalized almost complex structure} is an endomorphism 
$\kal{J}: T\oplus T^*\to T\oplus T^*$ that satisfies 
$\kal{J}^2=-1_{2d}$ and preserves  the natural metric $\kal{I}$,
$\kal{J}^t\kal{I}\kal{J}=\kal{I}$. 
The projection operators $\Pi_{\pm}\equiv \frac 1 2 (1\pm i\kal{J})$ 
are then used to define integrability (making $\kal{J}$ a generalized complex structure) as
\beq
\Pi_{\mp}[\Pi_{\pm}(X+\xi),\Pi_{\pm}(Y+\eta)]_{c}=0
\eeq{int2}

In a coordinate basis $\kal{J}$ is representable as
\beq
\kal{J}=\left(\begin{array}{cc}
J&P\cr
L&K\end{array}\right)~,
\eeq{Jcor1}
where $J:T\to T,\quad P:T^*\to T,\quad L:T\to T^*,\quad K:T^*\to T^*$. The condition 
$\kal{J}^2=-1_{2d}$ will impose the conditions 
\ber
&&J^2+PL=-1_{d}\cr
&&JP+PK=0\cr
&&KL+LJ=0\cr
&&LP+K^2=-1_{d}~,
\eer{cond3}
hermiticity of $\kal{I}$ implies
\ber
&&K=-J^t\cr
&&P^t=-P\cr
&&L^t=-L
\eer{pres}
and (\ref{int2}) will impose differential conditions on $J,P,L$ and $K$ (For their explicit form, 
see \cite{Lindstrom:2004iw}). 

An ordinary complex structure $J$ corresponds to the GCS
\beq
\kal{J}_{J}=\left(\begin{array}{cc}
J&0\cr
0&-J^t\end{array}\right)~,
\eeq{JJ}
and a symplectic structure $\omega$ corresponds to\footnote{For a generalized complex structure to
exist $T$ has to be even-dimensional.}
\beq
\kal{J}_{\omega}=\left(\begin{array}{cc}
0&-\omega^{-1}\cr
\omega&0\end{array}\right)~.
\eeq{Jo}
A $b$-transform acts as follows
\beq
\kal{J}_{b}=\left(\begin{array}{cc}
1&0\cr
b&1\end{array}\right)\kal{J}\left(\begin{array}{cc}
1&0\cr
-b&1\end{array}\right).
\eeq{Jcor2}

The general situation is illustrated in the following diagram:
\begin{center}
\includegraphics[width=8cm]{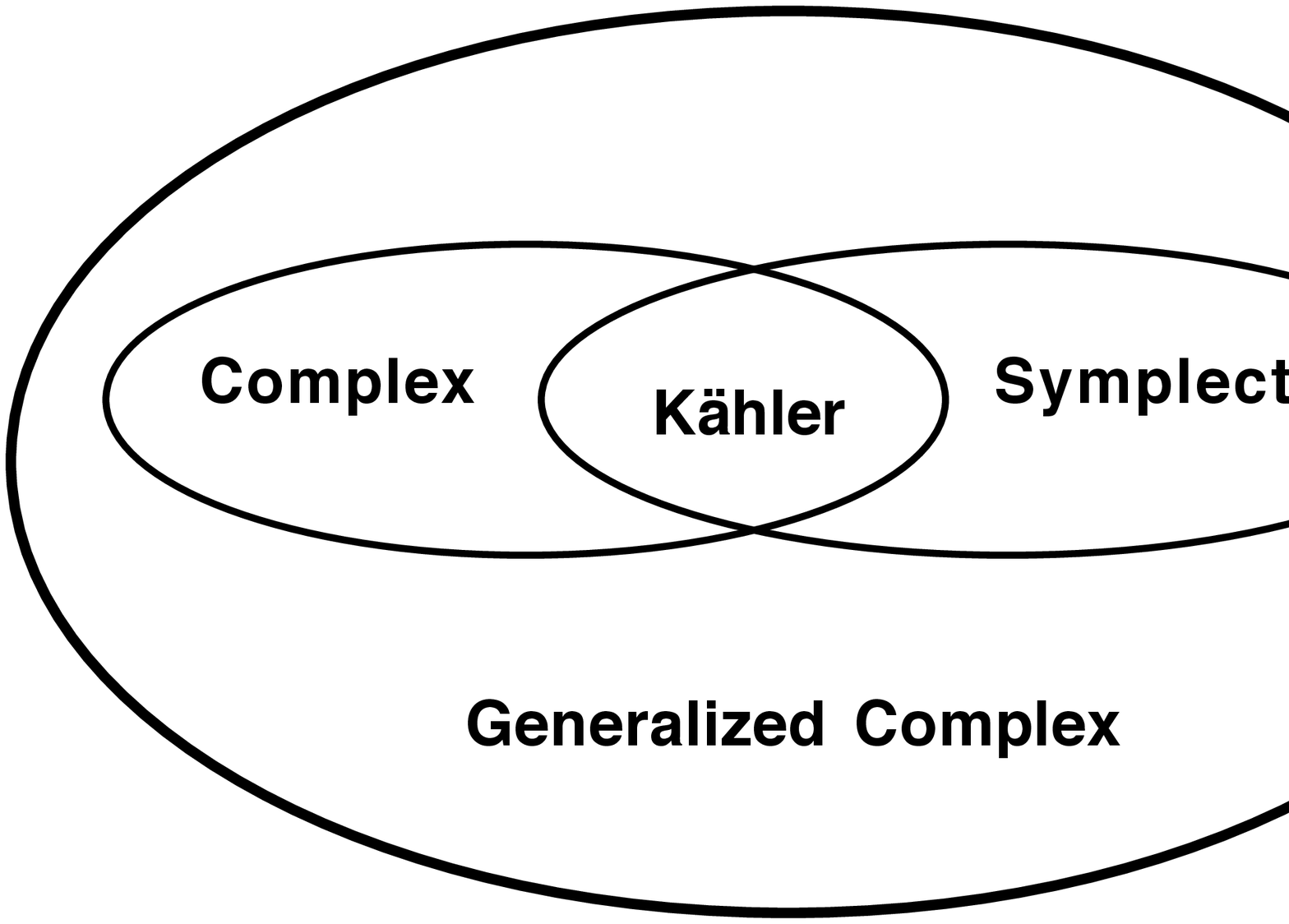}

Figure 1. The relation between the different geometries discussed.
\end{center}
A useful property for calculations is that locally (in an open set around a regular point) 
a manifold which 
admits a generalized complex structure may be brought to look like an open set in $\mathbb{C}^k$ 
times an open set in $(R^{2d-2k},\omega)$, where $\omega $ is in Darboux coordinates and $\mathbb{C}^k$
in complex (holomorphic and antiholomorphic) coordinates 
(using diffeomorphisms and $b$-transforms)\footnote{The proof of this, 
generalizing the Newlander-Nirenberg and the Darboux theorems, may be found in 
Gualtieri's thesis \cite{Gualtieri}, section 4.7.}.
\begin{center}
\includegraphics[width=8cm]{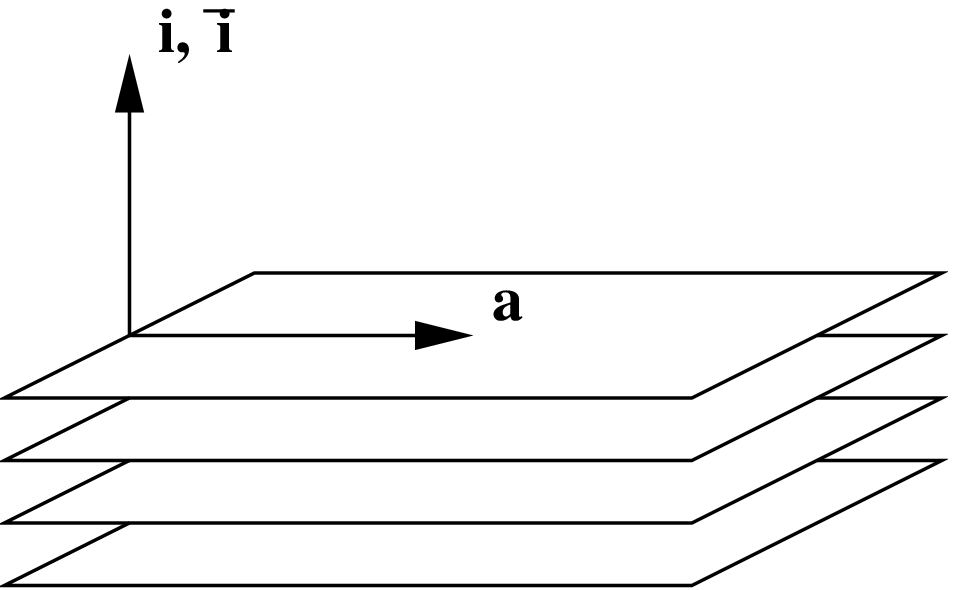}

Figure 2. A naive picture  of the local foliation in complex coordinates $z^{i}, \bar z^{i}$\\
and Darboux coordinates
$x^{a}$.
\end{center}

The generalized complex geometry is said to be {\em generalized K\"ahler} \cite{Gualtieri} 
if there exist\footnote{See chapter 6 of \cite{Gualtieri} for a full discussion of generalized \ka geometry and the relation to bihermitean geometry.} two commuting generalized 
complex structures $\kal{J}_{1}$ and $\kal{J}_{2}$ such that $\kal{G}=-\kal{J}_{1}\kal{J}_{2}$ is a positive definite 
metric on $T\oplus T^*$. For a  K\"ahler manifold $(J,g,\omega)$, using (\ref{JJ}) and (\ref{Jo}) one finds
the metric
\beq
\kal{G}=-\kal{J}_{J}\kal{J}_{\omega}=\left(\begin{array}{cc}
0&g^{-1}\cr
g&0\end{array}\right)~.
\eeq{Jcor}

One of the main results in \cite{Gualtieri} is that the bihermitean geometry with a $B$-field discovered in \cite{Gates:nk} is equivalent to
generalized \ka geometry.

Finally, it is worth mentioning that it is possible to twist the above structure by a closed three-form.

We now turn to the question of how this geometry may be realized in sigma-models.

\section{Sigma model realization}

In this section we recapitulate the basic set up used to discuss the relation between supersymmetric sigma-models and GCS's
in \cite{Lindstrom:2004iw},\cite{Lindstrom:2004eh}.

To realize the GCS in a sigma model, we need a formulation with additional fields 
$S\in T^*$. We thus consider the following first order action
\beq
S=\int d^2\xi d^2\theta\left(S_{\mu +}E^{\mu\nu}(\phi)S_{\nu -}-S_{\mu (+}D_{-)}\phi^\mu +
D_{+}\phi^\mu (B-b)_{\mu\nu}D_{-}\phi^\nu\right)~,
\eeq{suac1}
where $E_{\mu\nu}\equiv g_{\mu\nu}+ b_{\mu\nu}$ and its inverse may be thought of as  open string data:
\ber
E^{(\mu\nu)}=G^{\mu\nu}~,\quad E^{[\mu\nu]}=\theta^{\mu\nu}~.
\eer{opd}
In ({\ref{suac1}), $S_{\mu\pm}$ acts as an auxiliary field that extends 
the model to a sigma model on $T\oplus T^*$, and $b$ is a globally defined two-form that makes
it possible to display the $b$-transform. (Note that the original model ({\ref{suac1}) 
depends only on $H=dB$, and $B$ is thus typically only locally defined). Eliminating  
$S_{\mu\pm}$ we recover the action in ({\ref{sac2}). The $b$-transform is 
the statement that if in two actions of the form ({\ref{suac1}), $E_{\mu\nu}$ and $\tilde E_{\mu\nu}$ differ by a closed two-form $\tilde b$, the two actions are equivalent. 

The action ({\ref{suac1}) has many interesting limits. 
For example, if the metric is set to zero, the action 
describes a supersymmetric version of a Poisson sigma model \cite{Schaller:1994es}. In what follows 
we are not interested in the difference between the $B$
and $b$-fields, but set them equal each other, so the $\kal{N}=(1,1)$ action we study is
\beq
S=\int d^2\xi d^2\theta\left(S_{\mu +}E^{\mu\nu}(\phi)S_{\nu -}-S_{\mu (+}D_{-)}\phi^\mu \right)~,
\eeq{suac2}
The form of ansatz for the second 
supersymmetry ($\delta =\delta^{(+)}+\delta^{(-)}$) is determined by a dimensional analysis to be  
\cite{Lindstrom:2004iw} 
\ber
\delta^{(\pm)}\phi^\mu &=& \e^{\pm}\L^{A}_{\pm}A^{(\pm)\mu}_{A}\cr
\delta^{(\pm)}S_{\mu \pm} &=& \e^{\pm}\left(D_{\pm}\L^{A}_{\pm}B^{(\pm)}_{\mu A}
+\L^{A}_{\pm}\L^{B}_{\pm}C^{(\pm)}_{\mu AB}\right) {\vbox{\hsize=0pt \vskip 20pt}}\cr
\delta^{(\pm)}S_{\mu \mp} &=& \e^{\pm}\left(D_{\pm}\L^{A}_{\mp}M^{(\pm)}_{\mu A}
+D_{\mp}\L^{A}_{\pm}N^{(\pm)}_{\mu A}
+\L^{A}_{\pm}\L^{B}_{\mp}X^{(\pm)}_{\mu AB}\right)~, {\vbox{\hsize=0pt \vskip 20pt}}
\eer{gsusy2}
where $\L^{A}_{\pm}\equiv (D_{\pm}\phi^\mu ,S_{\mu \pm})$ lie in
$T\oplus T^*$, and all the coefficients
$A^{(\pm)\mu}_{A},B^{(\pm)}_{\mu A},C^{(\pm)}_{\mu AB},M^{(\pm)}_{\mu A},N^{(\pm)}_{\mu A},$ and
$X^{(\pm)}_{\mu AB}$ are functions of $\phi$. The conditions that follow from invariance of the action
and closure of the algebra are of two kinds: algebraic and differential. 
The two-index coefficients ($C$ and $X$) typically turn out to be given as derivatives
of the one-index coefficients ($A,B,M$ and $N$), just as the generalized complex structures  
are given in terms of $J,P,L$ and $K$, which subsequently obey differential conditions via the integrability requirement. In \cite{Lindstrom:2004iw} it is shown 
that the analogous index coefficients  for certain 
$\kal{N}=(2,0)$ sigma-models have a direct interpretation as submatrices of a GCS.

\section{Semi-Chiral Multiplets}

Semi-(anti)chiral superfields are left or right  chiral $\kal{N}=(2,2))$ multiplets and were
introduced in \cite{Buscher:1987uw}. They are represented by $\kal{N}=(2,2)$ superfields that 
obey ``half'' the usual chirality constraints.
A complex left chiral superfield $\mathbb{X}$ obeys
\beq
\bar \mathbb{D}_{+}\mathbb{X}=0~,
\eeq{left}
and a right antichiral superfield $\mathbb{Y}$ obeys
\beq
\mathbb{D}_{-}\mathbb{Y}=0~. 
\eeq{right}
Decomposing the $\kal{N}=(2,2)$ covariant 
derivatives $(\mathbb{D}_{\pm}, \bar \mathbb{D}_{\pm})$  into $\kal{N}=(1,1)$ derivatives $D_{\pm}$ and extra supercharges 
$Q_{\pm}$,
\ber
&&D_{\pm}=\mathbb{D}_{\pm}+\bar \mathbb{D}_{\pm}~,\cr
&&Q_{\pm}=i(\mathbb{D}_{\pm}-\bar \mathbb{D}_{\pm})~,
\eer{twone}
and the  $\kal{N}=(2,2)$ superfields into  $\kal{N}=(1,1)$ superfields, 
\ber
&&\varphi\equiv \mathbb{X}|\qquad \F_{-}\equiv Q_{-}\mathbb{X} |\cr
&&\chi \equiv \mathbb{Y}|\qquad \Upsilon _{+}\equiv Q_{+}\mathbb{Y} |~,
\eer{twone2}
the conditions (\ref{left}, \ref{right}) become 
\beq
\begin{array}{llll}
Q_{+}\varphi=iD_{+}\varphi ~,~~& 
Q_{-}\varphi=\F_{-} ~,~~& 
Q_{+}\F_{-}=iD_{+}\F_{-} ~,~~& 
Q_{-}\F_{-}=-i\d_{=}\varphi~,~~\\
Q_{+}\bar\varphi=-iD_{+}\bar\varphi ~,~~& 
Q_{-}\bar\varphi=\bar\F_{-} ~,~~& 
Q_{+}\bar\F_{-}=-iD_{+}\bar\F_{-} ~,~~& 
Q_{-}\bar\F_{-}=-i\d_{=}\bar\varphi~,~~\\
Q_{-}\chi=-iD_{-}\chi ~,~~& 
Q_{+}\chi=\Upsilon_{+} ~,~~& 
Q_{-}\Upsilon_{+}=-iD_{-}\Upsilon_{+} ~,~~& 
Q_{+}\Upsilon_{+}=-i\d_{\+}\chi~,~~\\
Q_{-}\bar\chi=iD_{-}\bar\chi ~,~~& 
Q_{+}\bar\chi=\bar\Upsilon_{+} ~,~~& 
Q_{-}\bar\Upsilon_{+}=iD_{-}\bar\Upsilon_{+} ~,~~& 
Q_{+}\bar\Upsilon_{+}=-i\d_{\+}\bar\chi~, \end{array}
\eeq{leftright}
where vector indices are denoted by pairs of spinor indices:  $\{\+,=\}$.

For $p$ left-chiral superfields $\mathbb{X}^{a},\, a=1,...,p$ and $p'$ right antichiral superfields 
$\mathbb{Y}^{a'}\!,\, a'=1,...,p'$, it is convenient to introduce the notation\footnote{Not to be confused with the previous usage of the index $A$ in section 4.}¥ $A=a,\bar a$ and $A'=a',\bar a'$. In this notation, we define the complex structures on the subspaces
\ber
&J^{A}_{~B}\equiv\left(\begin{array}{cc}
i\delta_{a}^{b}&0\cr
0&-i\delta_{\bar a}^{\bar b}\end{array}\right)&J^{A'}_{~B'}\equiv\left(\begin{array}{cc}
i\delta_{a'}^{b'}&0\cr
0&-i\delta_{\bar a'}^{\bar b'}\end{array}\right)~,
\eer{CS}
the relations \enr{leftright} can be written as
\beq
\begin{array}{llll}
Q_{+}\varphi^{A}=J^{A}_{~B}D_{+}\varphi^{B}~,&  Q_{+}\F_{-}^{A}=J^{A}_{~B}D_{+}\F_{-}^{B}~,
&   Q_{+}\chi^{A'}=\Upsilon_{+}^{A'},~&   Q_{+}\Upsilon_{+}^{A'}=-i\d_{\+}\chi^{A'},\\ 
Q_{-}\chi^{A'}=-J_{~B'}^{A'}D_{-}\chi^{B'},  &   Q_{-}\Upsilon_{+}^{A'}=
-J_{~B'}^{A'}D_{-}\Upsilon^{B'}, &  Q_{-}\varphi^{A}=\F_{-}^{A}~,&  Q_{-}\F_{-}^{A}=-i\d_{=}\varphi^{A}~. 
{\vbox{\hsize=0pt \vskip 20pt}}
\end{array}
\eeq{COV}
Despite the similarities with the geometrical discussion in 
Section 3 we cannot identify the 
generalized K\"ahler geometry of our model directly from \enr{COV}. 
If we restrict ourselves to $Q_+$ transformations, then the $A$-directions 
correspond to a transverse complex structure and the $A'$-directions to  
symplectic directions, thus indicating the relation to the generalized complex 
geometry. The same can be done for the $Q_-$ transformations. However, an
off-shell formulation never contains the full information about the geometry
({\it e.g.}, the metric), and thus an analysis of possible actions and their
invariances is needed.

\section{Actions}
\subsection{A topological model} 
A general action that depends only on $\mathbb{X}$ and $\bar\mathbb{X}$ gives a topological example:
\beq
S_{\mathbb{X}}=\int d^2\xi d^2\theta d^2\bar\theta K(\mathbb{X}^{a},\bar\mathbb{X}^{\bar a})
=\int d^2\xi \mathbb{D} ^2\bar \mathbb{D}^2 K(\mathbb{X}^{a},\bar\mathbb{X}^{\bar a})~.
\eeq{Act5}
Reducing to $\kal{N}=(1,1)$ using the results in section 5 in the standard fashion (see e.g. \cite{Gates:nk}), we find
\beq
S_{\mathbb{X}}= -\frac i 4 \int d^2\xi D^2 (\omega_{AB}D_{+}\varphi^{A}
\Psi^{B}_{-})=-\frac i 4 \int d^2\xi D^2( D_{+}\varphi^{A}S_{A-})~,
\eeq{Act6}
where we have redefined \cite{Buscher:1987uw}
\beq
S_{A-}=\omega_{AB}\F^{B}_{-}~,
\eeq{Psredef}
\bigskip
and introduced the symplectic form
\ber
&\omega_{AB}\equiv\left(\begin{array}{cc}
0&iK_{a\bar b}\cr
-iK_{\bar a b}&0\end{array}\right)~.
\eer{HM}
\subsection{Sigma models}
To construct a sigma model, we need to introduce an equal number of semi-chiral 
and semi-antichiral fields. This ensures that all the auxiliary superfields can be eliminated.
The most general manifest  $\kal{N}=(2,2)$ action with $p$ 
left chiral multiplets $\mathbb{X}^{a}$ and $p=p'$ right chiral multiplets $\mathbb{Y}^{a'}$ is
\beq
\int d^2\xi d^2\theta d^2\bar\theta K(\mathbb{X}^{A},\mathbb{Y}^{A'})
=\int d^2\xi \mathbb{D} ^2\bar \mathbb{D}^2 K(\mathbb{X}^{A},\mathbb{Y}^{A'})=-\frac i 4 \int d^2\xi D^{2}Q_{+}Q_{-}
K(\mathbb{X}^{A},\mathbb{Y}^{A'})
\eeq{Act2}
We reduce to $\kal{N}=(1,1)$ by acting with $Q_{\pm}$ and keeping the part independent of the second $\theta$.
The result is
\ber
S=\frac14 \int d^2\xi&\!\!\!\!D^2 &\!\!\!\! \left(-\,iD_{+}\varphi^{A}m_{AA'}D_{-}\chi^{A'}
-i\Upsilon^{A'}_{+}n_{A'A}\F^{A}_{-}\right.\cr 
&&\left.+\,\F^{A}_{-}(-2i\omega_{AB}D_{+}\varphi^{B}+p_{AB'}D_{+}\chi^{B'})\right.
{\vbox{\hsize=0pt \vskip 18pt}}\cr
&&\left.-\,\Upsilon^{A'}_{+}(q_{A'B}D_{-}\varphi^{B}+2i\omega_{A'B'}D_{-}\chi^{B'})\right)~,
{\vbox{\hsize=0pt \vskip 20pt}}
\eer{ACt6}
where the $2p\times 2p$ matrices are
\beq
\begin{array}{ccccc}
m_{AA'}&=&J^{B}_{~A}K_{B'B}J^{B'}_{~A'}&=&\left(\begin{array}{cc}
K_{aa'}&-K_{a\bar a'}\cr 
-K_{\bar a a'}&K_{\bar a\bar a'}\end{array}\right)\cr 
n_{A'A}&=&K_{A'A}&=&\left(\begin{array}{cc}
K_{a'a}&K_{a'\bar a}\cr
K_{\bar a'a}&K_{\bar a'\bar a}\end{array}\right) {\vbox{\hsize=0pt \vskip 30pt}}\cr
-2i\omega_{AB}&=& -iJ^{C}_{~[A}K_{B]C}&=&\left(\begin{array}{cc}
0 & 2K_{a\bar b}\cr
-2K_{a\bar b}&0\end{array}\right){\vbox{\hsize=0pt \vskip 30pt}}\cr
-2i\omega_{A'B'}&=& -iJ^{C'}_{~[A'}K_{B']C'}&=&\left(\begin{array}{cc}
0 & 2K_{a'\bar b'}\cr
-2K_{a'\bar b'}&0\end{array}\right){\vbox{\hsize=0pt \vskip 30pt}}\cr
p_{AA'}&=&-iJ^{C}_{~A}K_{CA'}&=&\left(\begin{array}{cc}
K_{aa'}&K_{a\bar a'}\cr
-K_{\bar a a'}&-K_{\bar a\bar a'}\end{array}\right){\vbox{\hsize=0pt \vskip 30pt}}\cr
q_{A'A}&=&iJ^{C'}_{~A'}K_{C'A}&=&~\,\left(\begin{array}{cc}
-K_{a'a}&-K_{a'\bar a}\cr
K_{\bar a'a}&K_{\bar a' \bar a}\end{array}\right)~.{\vbox{\hsize=0pt \vskip 30pt}}
\end{array}
\eeq{coeffmat}
Here $\omega_{AB}$ and $\omega_{A'B'}$ are symplectic structures.\\

Since \enr{ACt6} is not directly comparable to the action \enr{suac1} that was previously used to 
discuss relations to GCG, 
we now show how to relate them. First we rewrite \enr{ACt6} as\footnote{Here $t$ denotes transpose.}
\beq
S= \frac 1 4 \int d^2\xi D^2 \left({\bf \Psi_{+}^t }\mathbb{N}{\bf \Psi_{-}}
+{\bf \Psi_{+}^t}\mathbb{Q}D_{-}{\bf \Phi}
+{\bf \Psi_{-}^t}\mathbb{P}D_{+}{\bf \Phi}
+D_{+}{\bf \Phi ^{t}}\mathbb{M}D_{-}{\bf \Phi}\right)~,
\eeq{Act3}
where we have defined the $4p$ dimensional column vectors 
\beq
{\bf \Psi_{\pm}}\equiv \left(\begin{array}{c}\Psi_{\pm}^A\cr
\Upsilon_{\pm}^{A'}
\end{array}\right)~,\quad {\bf \Phi}\equiv 
\left(\begin{array}{c}\varphi^A\cr
\chi^{A'}
\end{array}\right)~,
\eeq{Vec}
and the $4p\times 4p$ matrices
\ber
&{\mathbb{N}\equiv -i\left(\begin{array}{cc}
0&0\cr
n&0
\end{array}\right)~,
\quad \mathbb{Q}\equiv -\left(\begin{array}{cc}
0&0\cr
q&2i\omega'\cr
\end{array}\right)}&,\cr
&&\,\cr
&&\,\cr
&{\mathbb{M}\equiv -i\left(\begin{array}{cc}
0&m\cr
0&0
\end{array}\right)}~,
\quad{\mathbb{P}\equiv \left(\begin{array}{cc}
-2i\omega&p\cr
0&0\cr\end{array}\right)}&~.
\eer{Mat} 
In the definition \enr{Vec}, we have introduced two new sets of fields $\F_{+}^{a}$ and $\Upsilon_{-}^{a'}$. 
From the form of the matrices in \enr{Mat} it follows that they do not appear in the action \enr{Act3}, but we shall need 
them below in discussing the relation to the action \enr{suac2}. \\

To compare (\ref{Act3}) to the action (\ref{suac2}) we  redefine the 
auxiliary fields ${\bf \Psi_{\pm}}$ in two steps. First we shift
\beq
{\bf \Psi_{\pm}}={\bf \hat \Psi_{\pm}}+\mathbb{A}^{(\pm)}D_{\pm}{\bf \Phi}~.
\eeq{shift}
This changes the Lagrangian in (\ref{Act3}) into
\ber
L &=& {\bf \hat\Psi_{+}^t} \mathbb{N}{\bf \hat\Psi_{-}}
+{\bf \hat\Psi_{+}^t}
(\mathbb{Q}+\mathbb{N}\mathbb{A}^{(-)})
D_{-}{\bf \Phi}
+{\bf \hat\Psi_{-}^t}
(\mathbb{P}-\mathbb{N}^t\mathbb{A}^{(+)})
D_{+}{\bf \Phi}\cr
&& +\,D_{+}{\bf \Phi^{t}}
\left(\mathbb{M}+\mathbb{A}^{(+)t }\mathbb{N}\mathbb{A}^{(-)}
-\mathbb{P}^t\mathbb{A}^{(-)}+\mathbb{A}^{(+)t}\mathbb{Q}\right)
D_{-}{\bf \Phi}~.{\vbox{\hsize=0pt \vskip 20pt}}
\eer{Lshift}
Next, to compare to \enr{suac2}, we want to choose the matices $\mathbb{A}^{(\pm)}$ to remove the last 
term in (\ref{Lshift}). The form of the matrices \enr{Mat} makes this impossible, however. 
This can be traced back to the properties of the left and right chiral multiplets we use; these contain only
the plus  or the minus auxiliary $\kal{N}=(1,1)$ spinor superfields. Thus the number of auxiliary fields 
does not match the number of coordinate fields. There are two ways to proceed: Either we compare to 
the form of the action that results from \enr{suac2} after integrating out some of the $S$-fields
or we ``reinstate'' the missing auxiliaries at the  $\kal{N}=(1,1)$ level by adding a 
trivial Lagrangian of the type $\kal{L}_{Extra}=\F_{+}^{a}\delta_{aa'}\Upsilon_{-}^{a'}$ in \enr{Act3}.  
It is for this purpose the fields $\F_{+}^{a}$ and $\Upsilon_{-}^{a'}$ were included in \enr{Vec}. Their 
supersymmetry transformations are simply
\ber
&Q_{+}\F_{+}^{a}=iD_{+}\F_{+}^{a}~,  &Q_{+}\Upsilon_{-}^{a'}=iD_{+}
\Upsilon_{-}^{a'}~,\cr
&Q_{-}\F_{+}^{a}=iD_{-}\F_{+}^{a}~,  &Q_{-}\Upsilon_{-}^{a'}=iD_{-}
\Upsilon_{-}^{a'}~,{\vbox{\hsize=0pt \vskip 23pt}}
\eer{stf}
and their algebra closes off-shell. We first discuss this case.

The effect of $\kal{L}_{Extra}$ is to change $\mathbb{N}\to\widetilde{\mathbb{N}}$ in \enr{Lshift}, where
\beq
\widetilde{\mathbb{N}}\equiv -i\left(\begin{array}{cc}
0&1\cr
m&0
\end{array}\right)
\eeq{Newmat}
is now invertible.  We may thus choose $\mathbb{A}^{(\pm)}$ such that $\mathbb{M}+\mathbb{A}^{(+)t }
\widetilde{\mathbb{N}}\mathbb{A}^{(-)}
-\mathbb{P}^t\mathbb{A}^{(-)}+\mathbb{A}^{(+)t}\mathbb{Q}=0$. Next we transform
\ber
&&{\bf \widetilde\Psi_{+}} =\left(\mathbb{Q}+\widetilde{\mathbb{N}}\mathbb{A}^{(-)}\right)^t{\bf \hat\Psi_{+}}\cr
&&{\bf \widetilde\Psi_{-}}=\left(\mathbb{P}-\widetilde{\mathbb{N}}^t\mathbb{A}^{(+)}\right)^t{\bf \hat\Psi_{-}}~.{\vbox{\hsize=0pt \vskip 20pt}}
\eer{scale}
With these transformations, the action (\ref{Act3})is brought to the form
\beq
S=  \int d^2\xi d^2\theta\left ({\bf \widetilde\Psi_{+}^t} \mathbb{E}^{-1}{\bf \widetilde\Psi_{-}}
+{\bf \widetilde\Psi_{(+}^t}
D_{-)}{\bf \Phi}\right)~,
\eeq{Act4}
with
\beq
\mathbb{E}^{-1}\equiv \left(\mathbb{Q}+\widetilde{\mathbb{N}}\mathbb{A}^{(-)}\right)^{-1}
\widetilde{\mathbb{N}}\left[\left(\mathbb{P}-\widetilde{\mathbb{N}}^t\mathbb{A}^{(+)} 
\right)^{t}\right]^{-1} ~.
\eeq{M}
This is the action (\ref{suac2}) in complex coordinates.

From (\ref{M}), we may thus read off the expression for $E^{\mu\nu}$, whose inverse gives us the 
background metric and antisymmetric field. Clearly both of these are given in terms of derivatives of $K$.
The transformations \enr{gsusy2} are then found by applying \enr{leftright} and \enr{stf} to the fields ${\bf \Phi}$ and ${\bf \widetilde\Psi_{\pm}}$.

In the alternative approach where we don't add $\kal{L}_{Extra}$, we choose $\mathbb{A}^{(\pm)}$ to have some convenient form and then make a  redefinition as in \enr{scale}. Independent of our choice, 
there will always be a nonzero $D_{+}{\bf \Phi}D_{-}{\bf \Phi}$-term. To compare to \enr{suac2} we must integrate out some $S_{\pm}$  to generate such a term. 

Finally, integrating out $\widetilde{\bf \Psi}_{\pm}$ from \enr{Act4} or 
${\bf \Psi_{\pm}}$ from \enr{Act3}, we find the sum of the metric 
and antisymmetric $B$-field to be given in terms of the potential $K$ as
\beq
\mathbb{E}=\mathbb{M}+\mathbb{P}^t\widetilde{\mathbb{N}}^{-1}\mathbb{Q}
\eeq{E}
\section{Geometric interpretation}
In this section we show that the transformations \enr{leftright} and 
\enr{stf} have a simple interpretation 
in terms of the GCG described in section 3. As mentioned there, we 
expect to find a generalized \ka geometry.
\subsection{A topological model}
We first discuss the topological model \enr{Act6}. 
Although this is not a proper sigma model, we expect 
it to be related to GCG in a manner analogous to
the toy-model discussed in \cite{Lindstrom:2004iw}. 
The second supersymmetry transformations are, from \enr{COV} and \enr{Psredef},
\ber
\delta^{(+)}\varphi^{A}&=&\e^{+}D_{+}\varphi^{B}J^{A}_{~B}~,\cr
\delta^{(-)}\varphi^{A}&=&-\e^{-}\omega^{AB}S_{B-}{\vbox{\hsize=0pt \vskip 22pt}}~,\cr
\delta^{(+)}S_{A-}&=&-\e^+D_{+}S_{B-}J^{B}_{~A}
+\omega_{AB}(\e^{+}J^{B}_{~C}D_{+}\varphi^{E}-\delta_{C}^{B}\delta^{(+)}\varphi^{E})\partial_E(\omega^{CD})
S_{D-}~,{\vbox{\hsize=0pt \vskip 22pt}}\cr
\delta^{(-)}S_{A-}&=&-i\e^-\omega_{AB}\d_{=}\varphi^{B}+
\omega^{BD}\partial_C(\omega_{AB})\delta^{(-)}\varphi^{C}S_{D-}~,
{\vbox{\hsize=0pt \vskip 22pt}}
\eer{Tfs+-}
where $\omega^{AB}$ is the inverse matrix of $\omega_{AB}$.
Closure of the algebra and invariance of the action are not issues here;
they are satisfied by construction (and are easily checked). 
These transformations represent a special case of the transformations \enr{gsusy2} 
(in complex coordinates).
As discussed in \cite{Lindstrom:2004iw}, the lowest tensor terms in the transformations 
typically give the generalized complex structure 
$\kal{J}$ while the higher terms are derivatives of lower ones needed for closure 
of the algebra, and geometrically ensure that closure of the algebra
(and invariance of the action) leads to integrability of $\kal{J}$. Here we already 
know that the GCS's will be integrable, since we start from a 
manifest $\kal{N}=(2,2)$ theory. We thus focus on identifying the GCS's and 
verify that they represent a generalized \ka geometry.  We find that also here the lowest tensor terms
suffice to identify the GCS. We then verify that it has the expected properties described in \cite{Gualtieri}.
From the terms in \enr{Tfs+-} not containing derivatives of the metric we identify the GCS
\beq
\kal{J}_{J}=\left(\begin{array}{cc}
J&0\cr
0&-J^{t}\end{array}\right)~,
\eeq{J1}
(with $J$ as defined in \enr{CS}) corresponding to the $(+)$ transformations. 
Since $\kal{J}_{J}$ is constant, integrability is immediate.
The $(-)$-transformations instead define
\beq
\kal{J}_{\omega}=\left(\begin{array}{cc}
0&-\omega^{-1}\cr
\omega&0\end{array}
\right)~,
\eeq{GCS1}
where $\omega_{AB}$ is the  the K\"ahler form  
corresponding to the \ka metric $g=\omega J$, which in turn defines the metric $\kal{G}$ on $T\oplus T^*$;
\beq
\kal{G}=\left(\begin{array}{cc}
0&g^{-1}\cr
g&0\end{array}\right)~,
\eeq{G}

We have thus uncovered the simplest case of generalized \ka geometry\footnote{It represents the 
ordinary \ka geometry $(J,g,\omega)$.} as defined in \cite{Gualtieri}, and as a final 
check we verify \enr{Jcor}:
\ber
-\kal{J}_{J}\kal{J}_{\omega}=\left(\begin{array}{cc}
0&g^{-1}\cr
g&0\end{array}\right)=\kal{G}~.
\eer{Gver}

We now turn to the full model in the form of the action \enr{ACt6}.
\subsection{Sigma models}
Using the field-redefinitions
\ber
u^{AA'}S_{A'-}&\equiv & \F^{A}_{-}-i\,u^{AA'}(q_{A'B}D_{-}\varphi^{B}
+2i\omega_{A'B'}D_{-}\chi^{B'})\cr
u^{AA'}S_{A+}&\equiv &\Upsilon^{A'}_{+}
-i\,u^{AA'}(-2i\omega_{AB}D_{+}\varphi^{B}+p_{AB'}D_{+}\chi^{B'})~,
{\vbox{\hsize=0pt \vskip 20pt}}
\eer{shift2}
where $u^{AA'}$ is the inverse of $n_{A'A}$,
the action \enr{ACt6} reads
\ber
S&=&-\frac i 4 \int d^2\xi\, D^2\left(D_{+}\varphi^{A}
(m_{AA'}+4\omega_{BA} u^{BB'}\omega_{B'A'})D_{-}\chi^{A'}\right.\cr
&&\qquad\qquad\qquad -\,2iD_{+}\varphi^{A}\omega_{BA}u^{BB'}q_{B'C}D_{-}\varphi^{C}
+D_{+}\chi^{A'}p_{AA'}u^{AB'}q_{B'B}D_{-}\varphi^{B}\cr
&&\qquad\qquad\qquad\left. +\,2iD_{+}\chi^{A'}p_{AA'}u^{AB'}\omega_{B'C'}D_{-}\chi^{C'}
+S_{A+}u^{AA'}S_{A'-}\right)\cr
&\equiv& -\frac i 4 \int d^2\xi D^2 \left( D_{+}{\bf \Phi}^{t}\mathbb{E}D_{-}{\bf \Phi}+S_{A+}u^{AA'}S_{A'-}\right)~,
\eer{RAct6}
with $\mathbb{E}$ the sum of the metric and the $B$-field,
\beq
\mathbb{E}=\left(\begin{array}{cc}
2i\omega uq&m-4\omega u\omega'\cr
p^{t}uq&2ip^{t}u\omega'\end{array}\right)
\eeq{E2}
This is an explicit form of the relation \enr{E} above.

The second $(+)$
supersymmetry transformations in \enr{leftright} read, with dots representing $\varphi$ and $\chi$-derivative terms, 
\ber
&\delta^{(+)}\varphi^{A}&=~\e^{+}D_{+}\varphi^{B}J^{A}_{~B}\cr
&\delta^{(+)}\chi^{A'}&=~\e^{+}u^{AA'}\left(S_{A+} +2\omega_{AB}D_{+}\varphi^{B}
+ip_{AB'}D_{+}\chi^{B'}\right)\cr
&\delta^{(+)}S_{A+}&=~\e^{+}\!\left(ip_{AA'}u^{BA'}D_{+}S_{B+}+(n_{A'A}
-p_{AB'}u^{BB'}p_{BA'})D_{+}^{2}\chi^{A'}\right.\cr
&&\qquad~~\left. +\,2(\omega_{AB}J^{B}_{~C}+p_{AA'}u^{BA'}
\omega_{BC})D^{2}_{+}\varphi^{C}\right)+...\cr
&\delta^{(+)}S_{A'-}&=~\e^{+}\!\left(-i(n_{A'A}J^{A}_{~B}u^{BB'}q_{B'C}-q_{A'A}J^{A}_{~C}-4i\omega_{A'B'}u^{BB'}\omega_{BC})D_{-}D_{+}\varphi^{B}\right.\cr
&&\qquad~~-\,2(n_{A'A}J^{A}_{~B}u^{BB'}\omega_{B'C'}
+i\omega_{A'B'}u^{BB'}p_{BC'})D_{-}D_{+}\chi^{C'}\cr
&&\qquad~~\left.-\,2\omega_{A'B'}u^{BB'}D_{-}S_{B+}+n_{A'A}J^{A}_{~B}u^{BB'}D_{+}S_{B'-}\right)+...~.
\eer{NSec}
From this we read off the following $8p\times 8p$ GCS 
\beq
\kal{J}^{(+)}=\left(\begin{array}{cccc}
J&0&0&0\cr
2u^t\omega&iu^tp& u^t&0\cr
-2 (\omega J+ipu^t\omega)&-  (n-pu^tp)&-ipu^t&0\cr
i  (-nJu^tq+qJ+4i\omega' u^t \omega)&2  (nJu\omega'-i\omega'u^tp)&
-2\omega'u^t&nJu\end{array}\right)~.
\eeq{Gcs7} 
It may be verified that $(\kal{J}^{(+)})^2=-\pmb{1}$, independent of the actual form \enr{coeffmat} of the sub-matrices\footnote{Although the entries in the matrix \enr{Gcs7} 
may be considerably simplified using \enr{coeffmat}.}.  Finally,  the form of the action \enr{RAct6} shows that on $S$-shell, {\it i.e.}, eliminating $S_{\pm}$, we recover the  usual
second order action studied by Gates, Hull and Ro\v cek in \cite{Gates:nk}.  
We must thus also recover the corresponding second supersymmetry 
transformations
\beq
\delta ^{(\pm)}{\bf \Phi}=\mathbb{J}^{(\pm)}\e^\pm D_{\pm}{\bf \Phi}~,
\eeq{ghrtf}
where ${\bf \Phi}$ is defined in \enr{Vec}. 
Setting 
$S_{\pm}=0$ in \enr{NSec} and comparing we see that the upper left hand submatrix $\mathbb{J}^{(+)}$ in \enr{Gcs7} must be an (ordinary) 
complex structure. It is indeed straightforward to show, using  \enr{coeffmat}, that 
\beq
(\mathbb{J}^{(+)})^{2} =\left(\begin{array}{cc}
J&0\cr
2u^t\omega&iu^tp\cr
\end{array}\right)^{2} =
-\left(\begin{array}{cccc}
\delta^{a}_{b}&0&0&0\cr
0&\delta^{\bar a}_{\bar b}&0&0\cr
0&0&\delta^{a'}_{b'}&0\cr
0&0&0&\delta^{\bar a'}_{\bar b'}\cr
\end{array}\right)~.
\eeq{CS7}
Here it is  important to note that  this part of the GCS \enr{Gcs7} represents the full $\varphi$ and $\chi$ transformations, {\it i.e.}, 
no derivative corrections were left out in \enr{NSec} in the transformations of these fields.

We now repeat the preceeding analysis for the second $(-)$ supersymmetry. The transformations in \enr{leftright} read
\ber
&\delta^{(-)}\varphi^{A}&=~\e^{-}u^{AA'}\left(S_{A'-} +iq_{A'B}D_{-}\varphi^{B}
-2\omega_{A'B'}D_{-}\chi^{B'}\right)\cr
&\delta^{(-)}\chi^{A'}&=~-\e^{-}D_{-}\chi^{B'}J^{A'}_{~B'}\cr
&\delta^{(-)}S_{A+}&=~
\e^{-}\!\left(2(n_{A'A}J^{A'}_{~B'}u^{BB'}\omega_{BC}+i\omega_{AB}u^{BB'}q_{B'C})
D_{+}D_{-}\varphi^{C}\right.\cr
&&\qquad~~+\,(in_{A'A}J^{A'}_{~B'}u^{BB'}p_{BC'}-4\omega_{AB}u^{BB'}\omega_{B'C'}-ip_{AB'}J^{B'}_{~C'})D_{+}D_{-}\chi^{C'}\cr
&&\qquad~~\left.+\,\,2\omega_{AB}u^{BB'}D_{+}S_{B'-}-
n_{A'A}J^{A'}_{~B'}u^{BB'}D_{-}S_{B+}\right)+...\cr
&\delta^{(-)}S_{A'-}&=~\e^{-}\!\left((-n_{A'C}+q_{A'A}u^{AB'}q_{B'C})D^{2}_{-}\varphi^{C}+iq_{A'A}u^{AB'}D_{-}S_{B'-}\right.\cr
&&\qquad~~\left.+\,2(iq_{A'A}u^{AB'}\omega_{B'C'}-\omega_{A'B'}J^{B'}_{~C'})D_{-}^{2}\chi^{C'}\right)+...~.
\eer{NSec2}
From this we read off the following GCS 
\beq
\kal{J}^{(-)}=\left(\begin{array}{cccc}
iuq&-2u\omega'&0&- u\cr
0&-J'&0&0\cr
-2(n^{t}J'u^t\omega+i\omega uq)&-i(n^{t}J'u^tp+4i\omega u\omega'-pJ')&-n^{t}J'u^t
&2\omega u\cr
n-quq&-2 (iqu\omega' - \omega'J')&0&-iqu\end{array}\right)~.
\eeq{Gcs8}
It may be verified that $(\kal{J}^{(-)})^2=-\pmb{1}$. The two GCS commute, {\it i.e.},
$[\kal{J}^{(+)},\kal{J}^{(-)}]=0$, as expected for a generealized \ka geometry. 
Again, the explicit form \enr{coeffmat} of the matrices
is only needed for checking that $(\mathbb{J}^{(-)})^{2}=-{\bf 1}$, 
where the complex structure is given by the upper left submatrix in \enr{Gcs8}.
Note that although the generalized complex structures commute, the complex structures do not, $[\mathbb{J}^{(+)},\mathbb{J}^{(-)}]\ne 0$.

The product of the two GCS's is
\ber
-\kal{G}&=&\kal{J}^{(-)}\kal{J}^{(+)}\cr
&&\ ~\cr
&=&\left(\begin{array}{cccc}
iJ uq&-2J u\omega'&0&- J  u\cr
-2J' u^t \omega&-iJ' u^t p&- J'  u^t &0\cr
2i(-\omega J  uq+pJ' u^t \omega)&4\omega J  u\omega'+nJ'-pJ'  u^t p&ipJ' u^t 
&2\omega J  u\cr
4\omega'J' u^t \omega+nJ-qJ uq&2i (-qJ u\omega' +\omega'J' u^t p)&2\omega'J' u^t &-iqJ u\end{array}\right)~.\cr
&&\ ~\ ~
\eer{GGG} 
As a final check that we are indeed dealing with a generalized \ka geometry, we have verified that indeed $\kal{G}^{2}=1$ (independent of the explicit 
form \enr{coeffmat} of the submatrices).

Note that the form of $\kal{J}^{(\pm)¥}¥$ depends on the choice of auxiliary
fields in the action. Getting from our form of the 
GCS's to the form 
given in chapter 6 of \cite{Gualtieri} involves a redefinition of the auxiliary
fields.

\section{Discussion}
In this paper we have studied the generalized \ka geometry present in nonlinear
sigma models where the 
basic fields are $\kal{N}=(2,2)$ semi-chiral 
multiplets. The underlying bihermitiean geometry has noncommuting complex
structures which
shows that the geometrical situation is different from that in a sigma model
with chiral and twisted chiral multiplets 
where the bihermitian complex structures necessarily commute. The geometry is
governed by a potential
$K(\mathbb{X},\bar\mathbb{X},\mathbb{Y},\bar\mathbb{Y})$ and our results should
be useful in constructing
explicit models (corresponding to choices of $K$), {\it e.g.}, generalizing the
discussions in \cite{Hitchin:1986ea}. 
Our models also lend themselves to topological twisting. Presumably the
auxiliary fields then have an interpretation as 
Batalin-Vilkovisky antifields \cite{Zucchini:2004ta}.

One open question is whether a nonlinear sigma model with chiral, twisted and
semi-(anti)chiral fields covers the
most general situation described by Generalized Complex Geometry. 
In the traditional bi-hermitean setting in, {\it e.g.,} 
\cite{Sevrin:1996jq},\cite{Bogaerts:1999jc}, claims in both directions have been made.
We have not proven or disproven this; clearly
a geometric understanding of the conditions that allow a description of sigma models 
using semi-(anti)chiral multiplets would cast light on this issue.

Finally, it should be interesting to study the model discussed when including
boundaries. Investigating the possible boundary
conditions that preserve some supersymmetry in an open sigma model has proven a
very powerful tool in understanding the geometry
of the $D$-brane where it ends \cite{Lindstrom:2002jb},\cite{Lindstrom:2002vp},
and recently it has been shown that these
investigations benefit from being formulated in terms of GCG \cite{Zabzine:2004dp}.

\noindent{\bf Acknowledgement}:  We are grateful to Marco Gualtieri for very 
helpful comments and to the 2004 Simons Workshop
for providing the stimulating atmosphere where this work was initiated.
The research of U.L. is supported by VR grant 650-1998368.  
The research of R.v.U. was supported by Czech ministry of education contract 
No. 143100006 and by Kontakt grant ME649. The research of M.Z. was supported by EU-grant
MEIF-CT-2004-500267. The work of MR was supported in part by NSF grant no.~PHY-0354776}
and Supplement for International Cooperation with Central and Eastern Euorpe PHY 0300634.

\end{document}